\documentclass[10pt,a4paper,twoside]{article}
\usepackage{amsfonts,psfig,graphicx}
\usepackage{etc9macro}

\begin{document}

\begin{etcpaper}

\begin{etctitle}
  \etcsettitle{Reacting compressible turbulence}
    {Non-premixed Flame-Turbulence Interaction in Compressible Turbulent Flow}
  \etcaddauthor{1}{Livescu}{D.}
  \etcaddlastauthor{2}{Madnia}{C.K.}
  \etcaddaffiliation{1}{University of California, \break
                 Los Alamos National Laboratory, \break
                 T-3/MS B216, Los Alamos, NM 87545, USA.}
  \etcaddaffiliation{2}{Department of Mechanical and Aerospace Engineering, \break
                 State University of New York at Buffalo, \break
                 Buffalo, NY 14260, USA.}
  \etcemailaddress{madnia@buffalo.edu}
  \etcsetpaperid{194}
\end{etctitle}

\section{Introduction}

Nonpremixed turbulent reacting flows are intrinsically difficult to 
model due to the strong coupling between turbulent motions and 
reaction. The large amount of heat released by a typical hydrocarbon 
flame leads to significant modifications of the thermodynamic 
variables and the molecular transport coefficients and thus alters 
the fluid dynamics \cite{JLM00},\cite{VP98}. 
Additionally, in nonpremixed 
combustion, the flame has a complex spatial structure. Localized 
expansions and contractions occur, enhancing the dilatational 
motions. Therefore, the compressibility of the flow and the heat 
release are intimately related. However, fundamental studies of the 
role of compressibility on the scalar mixing and reaction are scarce.
In this paper we present results concerning the fundamental aspects of the 
interaction between non-premixed flame
and compressible turbulence.

\section{Results}

In order to assess the influence of compressibility on the coupling 
between turbulence and reaction, direct numerical simulations (DNS) of
decaying isotropic and homogeneous sheared turbulence are performed for
different initial values of the turbulent Mach number, $M_{t_0}$, under
reacting and nonreacting conditions. The continuity, 
momentum, energy and species mass fractions transport equations are 
solved using the spectral collocation method. 
The chemical reaction is modeled as one-step, exothermic, and Arrhenius type. 
The scalar fields are initialized with a double-delta PDF (``random blobs''). 
The range of non-dimensional mean shear rates considered for homogeneous shear flow cases extends 
from 4.8 to 22. In this paper the results for $S^*=7.24$, which
are in the range dominated by nonlinear effects, are presented.
 The reaction parameters mimic the low to moderate Reynolds number combustion of a typical 
hydrocarbon in air. Details about the numerical method as well as 
the influence of the reaction parameters on  the flame-turbulence 
interaction can be found in Livescu, Jaberi \& Madnia \cite{LJM02}.
The range of Mach numbers considered in the present study extends 
from a nearly incompressible case with $M_{t_0}=0.1$ to $M_{t_0}=0.6$ 
which is at the upper limit for the numerical method considered.
The decaying isotropic turbulence simulations have the value of the Taylor 
Reynolds number $Re_{\lambda_0}=55$ at the time when the scalar field is initialized.
For the turbulent shear flow simulations the range of $Re_{\lambda_0}$ extends from 21 to 50. In this paper the results corresponding to $Re_{\lambda_0}=21$ 
are considered. The results obtained for higher $Re_\lambda$ cases are 
consistent with those presented in this paper.  

Figure \ref{f1}(a) shows that for turbulent shear flow cases the 
peak of the mean reaction rate decreases its magnitude and occurs at 
earlier times as the value of $M_{t_0}$ is increased. Similar results 
are obtained for the isotropic turbulence cases, although the $M_{t_0}$ 
influence on the evolution of the reaction rate is weaker than in 
shear flow. In order to better understand this 
behavior, the reaction rate, $w=Da \rho^2 Y_A Y_B exp(-Ze/T)$, is 
decomposed into its components, the mixing term, $G=\rho^2 Y_A Y_B$, 
and the temperature dependent term, $F= exp(-Ze/T)$ \cite{JLM00}. 
For turbulent shear flow, the results presented in 
figure 1(b) indicate that the Mach number has a different influence
on the evolutions of $F$ and $G$.
As $M_{t_0}$ increases $F$ increases, indicating elevated temperatures.
 The increase in the temperature at
higher $M_{t_0}$ can be associated with an enhanced viscous dissipation 
in the mean temperature transport equation. Moreover, 
this effect is more important for turbulent shear flow
where the viscous dissipation levels are higher. For both flows 
considered, higher temperatures expedite the ignition and the mean reaction
rate peaks at an earlier time. However, at earlier times the reactants
are less mixed so that the mixing term $G$ has lower values. 
The combined effects of $F$ and $G$ terms result in a decrease in the 
magnitude of the peak of the mean reaction rate.

The results presented above are consistent with those of Livescu
\& Madnia \cite{LM02} which show that the scalar mixing is more 
sensitive to changes in Mach number in homogeneous turbulent shear 
flow than in isotropic turbulence. For turbulent shear flow, due
to the presence of a mean velocity gradient the scalar field 
develops a preferential spatial orientation. Figure \ref{f2}(a) shows 
that the angle between the direction $x_1$ of the mean velocity, 
and the scalar gradient projection on the plane formed by the 
direction $x_2$ of the shear and $x_1$, has a most probable 
distribution approaching values close to $\pm90^\circ$ after some development time. This 
indicates that the scalar blobs are distorted into parallel layers
oriented at a small angle with respect to the plane perpendicular to the
direction of the shear. 

\begin{figure}[!ht]
\begin{minipage}[c]{.95 \linewidth}
\vskip 1.5cm
\begin{center}
\includegraphics[width=\linewidth]{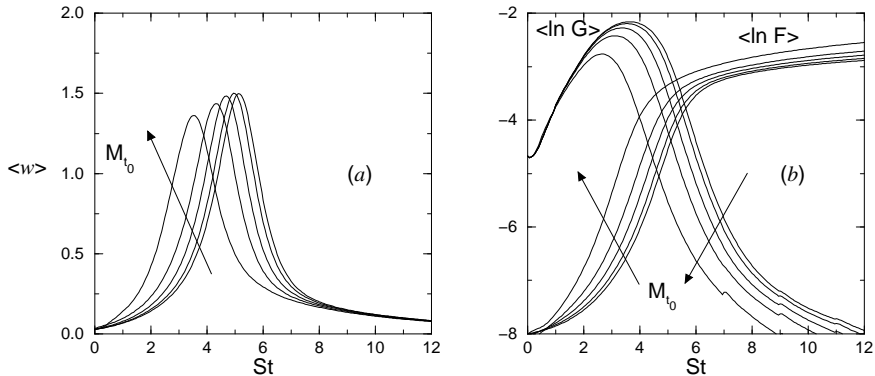}
\end{center}
\end{minipage}
\caption{Mach number influence on the evolution of (a) mean reaction
rate and b) reaction rate components in turbulent shear flow.}
\label{f1}
\end{figure}

Due to the spatial structure of the scalar field, the relative
orientation between the scalar gradients and different quantities 
pertaining to the velocity field are different in turbulent shear
flow than in isotropic turbulence. In particular, Livescu \& Madnia
\cite{LM02} show that a passive scalar gradient is no longer aligned 
with the most compressive eigenvector of the solenoidal strain rate
tensor as in isotropic turbulence and the relative angle changes with
Mach number. As a result, the production term in the scalar dissipation
equation decreases as $M_{t_0}$ is increased and the mixing becomes 
worse. Similar results are obtained for a reacting scalar. However,
in this case the heat of reaction affects the turbulence. Since the 
reaction takes place mostly at the interface between the scalar layers,
the localized expansions and contractions due to the heat of reaction 
also develop a preferential spatial orientation. 
 
For a nonreacting homogeneous shear flow it is known that the explicit
dilatational effects occur predominantly in the direction of the shear
\cite{LJM02}. In the reacting case, due to the anisotropy
in the heat release, these effects are further amplified. In 
particular, figure \ref{f2}(b) shows that the dilatational kinetic 
energy in the direction of the mean shear is increased by the reaction.
Moreover, this increase is more significant at higher values of 
$M_{t_0}$. 

\begin{figure}[!ht]
\begin{minipage}[c]{.95 \linewidth}
\vskip 1.5cm
\begin{center}
\includegraphics[width=\linewidth]{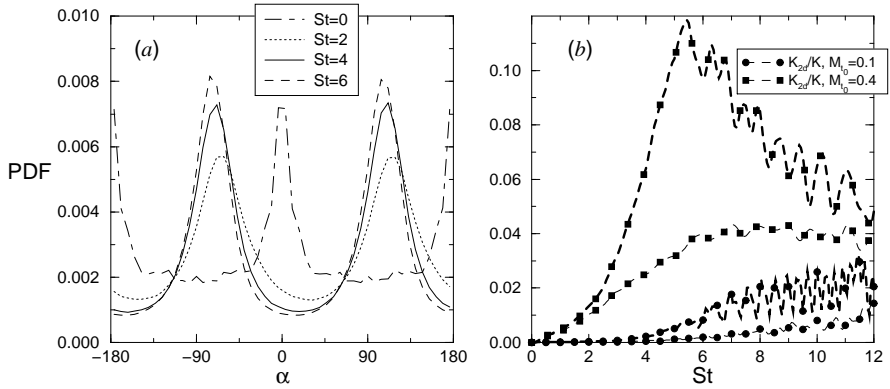}
\end{center}
\end{minipage}
\caption{a) PDF of the angle between the $x_1$ axis and the scalar 
gradient projection on $x_1-x_2$ plane at different times and b) 
Mach number effect on the dilatational kinetic energy in $x_2$ direction for 
nonreacting (thin lines) and reacting (thick lines) cases.}
\label{f2}
\end{figure}

\section{Concluding Remarks}

The Mach number effect on the two-way interaction between turbulence and
non-premixed flame is studied in isotropic turbulence and homogeneous turbulent 
shear flow using data generated by DNS. The results show that the
reaction rate and its components is less affected by changes in 
$M_{t_0}$ in isotropic turbulence than in turbulent shear flow. For
the latter case, the scalar field, and thus the reaction, develops a 
preferential spatial orientation. As a result, the relative orientation between
the scalar gradients and different quantities pertaining to the velocity field
is different than in isotropic turbulence. This affects the mixing process and 
leads to an increased sensitivity to the initial value of the turbulent Mach number.
Moreover, the anisotropy in the explicit dilatational effects is significantly
amplified by the reaction.

\end{etcpaper}

\begin{thebibliography}{1}

\bibitem{JLM00} 
F.~A. Jaberi, D.~Livescu, and C.~K. Madnia.
\newblock Characteristics of chemically reacting compressible homogeneous 
turbulence. 
\newblock {\em Physics of Fluids} 12:1189--1209, 2000. 

\bibitem{LM02} 
D.~Livescu, and C.~K. Madnia.
\newblock Compressibility effects on the scalar mixing in reacting 
homogeneous turbulence. 
\newblock In {\em Turbulent Mixing and Combustion}, Editors:
 A. Pollard and S. Candel, Kluwer Academic Press, in press. 

\bibitem{LJM02} 
D.~Livescu, F.~A. Jaberi, and C.~K. Madnia. 
\newblock The effects of heat release on the energy exchange in reacting 
turbulent shear flow. 
\newblock {\em J. Fluid Mech.}, 450:35--66, 2002. 

\bibitem{VP98} 
L.~Vervisch, and T.~Poinsot.
\newblock Direct numerical simulation of non-premixed turbulent flames. 
\newblock {\em Annu. Rev. Fluid Mech.} 30:655-691, 1998. 

\end{thebibliography}
\end{document}